\documentclass[conference]{IEEEtran}
\IEEEoverridecommandlockouts

\usepackage{graphicx}
\usepackage{xcolor}
\usepackage{multirow}
\usepackage{booktabs}
\usepackage{amsmath}
\usepackage{amssymb}
\usepackage{cite}
\usepackage{cuted}
\newcommand{\linebreakand}{%
  \end{@IEEEauthorhalign}
  \hfill\mbox{}\par
  \mbox{}\hfill\begin{@IEEEauthorhalign}
}    
\usepackage[colorlinks=Flase]{hyperref}
\usepackage{bbm}
\usepackage{algorithm} 
\usepackage{algpseudocode}
\usepackage{caption}
\usepackage{steinmetz}
\usepackage{subcaption}
\usepackage{amsthm}
\usepackage[english]{babel}


\newcounter{relctr} 
\everydisplay\expandafter{\the\everydisplay\setcounter{relctr}{0}} 
\usepackage[figurename=Fig.]{caption}
\newcommand{\norm}[1]{\left\lVert#1\right\rVert}
\usepackage{xpatch}
\newcommand{\C}{\mathbb{C}}

\newtheoremstyle{remarkstyle}%
  {}
  {}
  {\itshape}
  {}
  {\itshape}
  {.}
  {.5em}
  {}

\theoremstyle{remarkstyle}

\AtBeginDocument{} 

\ifCLASSINFOpdf
\else
\fi

\begin{document}

\title{Echo-Conditioned Denoising Diffusion Probabilistic Models for Multi-Target Tracking in RF Sensing\\
\thanks{This work is supported by the Research Council of Finland (Grants 362782 (ECO-LITE), and 369116 (6G Flagship)).}

\author{
{ Amirhossein~Azarbahram and Onel~L.~A.~L\'opez}
\vspace{1mm}
\\

\small Centre for Wireless Communications (CWC), University of Oulu, Finland \\
\small Emails: \{amirhossein.azarbahram, onel.alcarazlopez\}@oulu.fi}
}

\maketitle

\begin{abstract}

In this paper, we consider a dynamic radio frequency sensing system aiming to spatially track multiple targets over time. We develop a conditional denoising diffusion probabilistic model (C-DDPM)-assisted framework that learns the temporal evolution of target parameters by leveraging the noisy echo observations as conditioning features. The proposed framework integrates a variational autoencoder (VAE) for echo compression and utilizes classifier-free guidance to enhance conditional denoising. In each transmission block, VAE encodes the received echo into a latent representation that conditions DDPM to predict future target states, which are then used for codebook beam selection. Simulation results show that the proposed approach outperforms classical signal processing, filtering, and deep learning benchmarks. The C-DDPM-assisted framework achieves significantly lower estimation errors in both angle and distance tracking, demonstrating the potential of generative models for integrated sensing and communications.

\end{abstract}


\IEEEpeerreviewmaketitle

\section{Introduction}

\IEEEPARstart{N}{ext}-generation wireless systems are expected to embed sensing capabilities for detecting, localizing, and tracking surrounding objects \cite{ISAC_main_survey}. Such integrated sensing and communications (ISAC) convergence, driven by higher carrier frequencies, large antenna arrays, and advanced waveforms, paves the way for applications such as vehicular safety, industrial automation, smart infrastructure, and massive internet of things (IoT) localization and monitoring.

Accurate sensing in such systems is challenging due to the dynamic environments, with target positions, velocities, and reflectivities changing rapidly, causing time-varying channels and non-stationary echoes. To maintain situational awareness, sensing algorithms also include prediction and refinement stages across consecutive observations. Indeed, tracking the temporal evolution of targets, rather than performing independent per-frame estimation, is crucial for achieving consistent/robust sensing in dynamic scenes, despite noise and clutter. This work precisely focuses on tracking-oriented radio frequency (RF) sensing that reconstructs the spatial–temporal geometry of the environment from echo signals. 

Classical signal processing (SP) approaches, such as subspace-based methods like multiple signal classification (MUSIC) \cite{musicref} and estimation of signal parameters via rotational invariance techniques (ESPRIT) \cite{esprit_ref}, have been widely used for angle estimation. These techniques offer high resolution but tend to degrade under moderate SNR, limited snapshots, or strong multipath and clutter conditions. Deep learning (DL) methods can overcome these limitations \cite{CNN_ISAC}, but typically require large labeled datasets and generalize poorly in dynamic scenarios due to their regressive nature. Hence, there is a need for new approaches that explicitly model uncertainty and exploit limited, noisy observations, which is also a key issue in resource-constrained IoT sensing networks.

Generative artificial intelligence (GAI) offers such a paradigm. GAI supports denoising, augmentation, and predictive behavior by learning complex data distributions and producing realistic samples. Starting from autoencoders and probabilistic models, the field advanced through generative adversarial networks (GANs) and variational autoencoders (VAEs), and has recently been revolutionized by denoising diffusion probabilistic models (DDPMs), which achieve unprecedented sample quality and robustness \cite{GenAI_history}. This generative capability, already demonstrated in content creation in large-scale foundation models such as ChatGPT, opens opportunities for RF sensing and IoT data reconstruction where uncertainty is a challenge.

From a sensing perspective, GAI has demonstrated promising results in channel state information (CSI) compression, estimation, beamforming, and signal enhancement for ISAC systems \cite{ISAC_genAI_mag}, motivating recent efforts to further integrate GAI into ISAC design. For example, a two-stage diffusion-based augmentation framework generates and refines CSI samples to alleviate data scarcity in \cite{ISAC_GEN_Augnment}, while a diffusion-based secure sensing system introduces safeguarding signals against unauthorized inference in \cite{ISAC_GEN_Secure}. DDPMs are incorporated into digital twins of ISAC channels for CSI estimation and target detection in \cite{ISAC_GEN_digitalTwin}, and conditional GANs are used for reconfigurable intelligent surfaces-assisted CSI estimation in \cite{ISAC_GEN_CSI}. Yet, the application of GAI models to multi-target RF tracking remains unexplored. Unlike CSI estimation or single-frame sensing enhancement, tracking requires temporal reasoning over blocks to understand motion dynamics under uncertainty. Classical SP or learning-based predictors typically rely on explicit state-space models, e.g., Kalman \cite{kalman_main} or particle filters \cite{particle_main}, that assume simplified motion or noise statistics, making them unsuitable for cluttered and non-stationary environments. GAI models can instead learn the conditional distribution of future target states given noisy or partial observations, capturing the multi-modal and correlated evolution of targets. Such capability is crucial for robust ISAC operation in dynamic scenes where targets and background reflections evolve unpredictably. We precisely corroborate this in a dynamic RF sensing system that tracks multiple targets across transmission blocks, with the design goal of minimizing the estimation error of their angles and distances.

\begin{figure}
    \centering
    \includegraphics[width=\columnwidth]{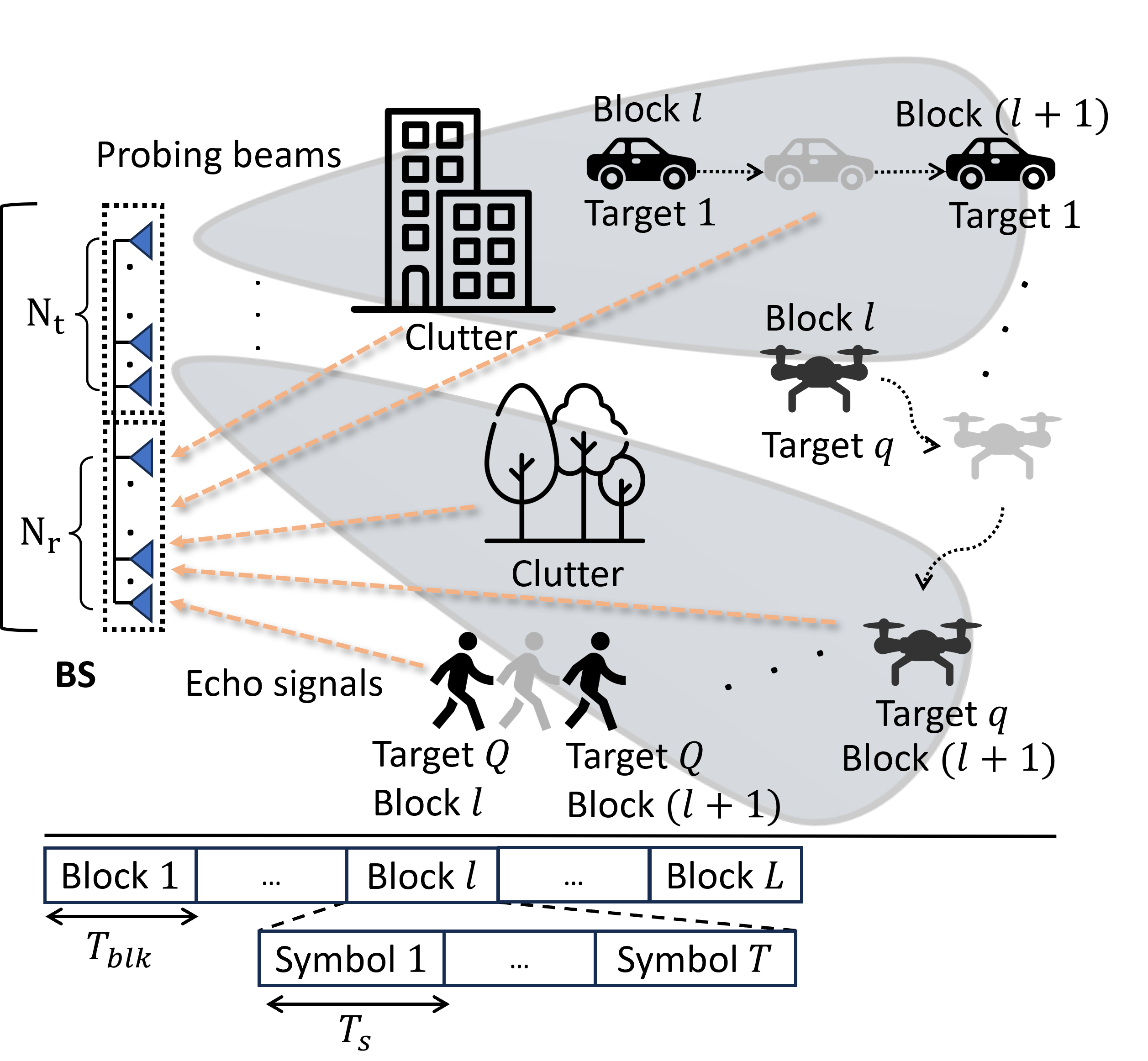}
    \caption{A RF sensing system, wherein BS transmits probing beams toward multiple targets and clutters, receives the reflected echo signals, and updates the target states across transmission blocks, each comprising $N$ sensing symbols.}
    \label{fig:sysmod_ddpm}
\end{figure}

\textbf{Contributions:} We develop a conditional DDPM (C-DDPM)-assisted framework for multi-target RF tracking that learns the temporal evolution of target parameters over successive blocks. The framework integrates a VAE for echo compression and classifier-free guidance to enhance conditional denoising. In each block, the VAE encodes the received echo into a latent representation that forms the conditioning vector for the C-DDPM, which then predicts the next-block target states. These predictions are used for codebook beam selection via the expected amplitude-weighted array gain. Simulation results demonstrate that the proposed framework consistently outperforms classical SP, filtering, and DL benchmarks for different numbers of targets. This confirms the ability of the proposed framework to learn temporal dynamics and uncertainty directly from echo observations, enabling robust and accurate multi-target tracking in ISAC settings.

\textbf{Notations:} Bold lower- and upper-case letters represent vectors and matrices, respectively. The $\ell_2$-norm operator is denoted by $\norm{\cdot}$. The Hermitian (conjugate transpose) is represented by $(\cdot)^H$, while $\Re\{\cdot\}$ and $\Im\{\cdot\}$ denote the real and imaginary parts, respectively. The symbol $\mathbf{I}_N$ denotes the $N{\times}N$ identity matrix. The notation $\mathcal{N}(\boldsymbol{\mu},\boldsymbol{\Sigma})$ represents a Gaussian distribution with mean vector $\boldsymbol{\mu}$ and covariance matrix $\boldsymbol{\Sigma}$; equivalently, $\mathcal{N}\!\big(\mathbf{x};\,\boldsymbol{\mu},\,\boldsymbol{\Sigma}\big)$ denotes the corresponding probability density function evaluated at $\mathbf{x}$. Finally, $\mathrm{Unif}\{a{:}b\}$ denotes a discrete uniform distribution over integers $a$ to $b$.

\section{System Model and Problem Formulation}\label{sec:sys}

We consider a narrowband sensing system with a full-duplex base station (BS) equipped with a uniform linear array (ULA) comprising $N_t$ transmit, and $N_r$ receive antennas. The carrier frequency is $f$, hence the wavelength is $\lambda = \tfrac{c}{f}$, and the inter-element spacing in the ULAs is $\lambda/2$. Time is partitioned into $L$ blocks, each comprising $N$ symbol slots, during which the system is assumed to remain static, i.e., target features (e.g., location and scattering) are fixed within a block but vary between blocks. The system model is illustrated in Fig.~\ref{fig:sysmod_ddpm}. The transmit and receive steering vectors to an impinging (or departing) signal from direction $\theta$ are given by
\begin{subequations}
\begin{align}
  \mathbf{a}_t(\theta) &= \frac{1}{\sqrt{N_t}}
  \big[\,1,\, e^{\jmath \pi \sin\theta},\, \cdots,\, e^{\jmath \pi (N_t-1)\sin\theta}\,\big]^{{T}}, \\
  \mathbf{a}_r(\theta) &= \frac{1}{\sqrt{N_r}}
  \big[\,1,\, e^{\jmath \pi \sin\theta},\, \cdots,\, e^{\jmath \pi (N_r-1)\sin\theta}\,\big]^{{T}}.
\end{align}
\end{subequations}
\subsection{Transmit Sensing and Received Echo Signals}

Each block~$l$ corresponds to a distinct sensing scene, characterized by potentially different target features and propagation conditions. In block~$l$ and slot~$n$, the BS transmits a superposition of $M_s \le N_t$ sensing beams
\begin{align}
\mathbf{s}_l[n] 
&= \sum_{m=1}^{M_s} \sqrt{P_{m,l}}\, \mathbf{v}_{m,l}\, e_{m,l}[n] \;\in\C^{N_t}, 
\end{align}
where $\sum_{m=1}^{M_s} P_{m,l} \le P_{\textrm{Tx}}$ is the per-block power constraint and $\{e_{m,l}[n]\}$ are normalized independent probing symbols. Each sensing beam $\mathbf{v}_{m,l}$ is selected from a finite transmit codebook $\mathcal{C}_\mathrm{s}=\{\mathbf{v}_1,\ldots,\mathbf{v}_{N_\mathrm{s}}\}$. We consider monostatic radar with $Q$ point targets, each characterized by an angle $\theta_{q,l}$, a distance to the BS $d_{q,l}$, Doppler frequency $f'_{q,l}=\tfrac{2v_{q,l}}{\lambda}$ with radial velocity $v_{q,l}$, and complex coefficient $\beta_{q,l}$. The coefficient captures both round-trip path-loss attenuation and radar cross-section (RCS). In addition, $P_l$ clutters with parameters $\gamma_{p,l}$ and Doppler frequency $\bar{f}'_{p,l}$ are present.
The received echo at slot $n$ is therefore given by \cite{ISAC_main_survey}
\begin{align}
\mathbf{r}_l[n]
&= \sum_{q=1}^{Q} \beta_{q,l}\, 
e^{j2\pi f_{q,l}' t_n}\,
\mathbf{a}_r(\theta_{q,l})\, 
\mathbf{a}_t(\theta_{q,l})^{\!H}\mathbf{s}_l[n] \label{eq:vector-echo-mono-dopp}\nonumber \\[-0.25em]
&\quad + \sum_{p=1}^{P_l} \gamma_{p,l}\, 
e^{j2\pi \bar{f}'_{p,l} t_n}\,
\mathbf{a}_r(\varphi_{p,l})\, 
\mathbf{a}_t(\varphi_{p,l})^{\!H}\mathbf{s}_l[n]
\;+\; \mathbf{z}_l[n], 
\end{align}
where $\mathbf{z}_l[n]\!\sim\!\mathcal{CN}(\mathbf{0},\sigma_r^2 \mathbf{I}_{N_r})$ is the additive white Gaussian noise, $t_n = (n - 1)T_s$, and $T_{\textrm{blk}} = N T_s$ is the block length.
Hereby, the concatenated received echo matrix at the end of $l$-th block is written as $\mathbf{R}_l \triangleq \big[\mathbf{r}_l[1],\dots,\mathbf{r}_l[N]\big] \in \C^{N_r\times T}.$

\subsection{Problem Formulation}

Classical ISAC designs typically formulate optimization objectives based on Cramer-Rao bound (CRB) \cite{CRB_opt}, beampattern shaping \cite{beampattern_trahertz}, or information-theoretic measures \cite{ISAC_information}. While helpful, such metrics are indirect, and our actual objective is to precisely estimate the target parameters needed for sensing and scheduling, namely the angles and distances. Thus, we formulate the per-target loss as
\begin{align}
\ell_{q, l}\!\left(\hat{\theta}_{q,l},\hat{d}_{q,l}\right)
= \, \!\left(\theta_{q,l}-\hat{\theta}_{q,l}\right)^2  
 +\, \eta\Big(d_{q,l}-\hat{d}_{q,l}\Big)^2,
\label{eq:per-target-loss-split}
\end{align}
where $\eta$ is the weight factor, while $\hat{\theta}_{q,l}$ and $\hat{d}_{q,l}$ are the estimated angle and distance, respectively. Note that angles and distances are directly observable from the echo model, while using a direct localization error would couple angular and range uncertainties nonlinearly, obscuring their contributions to sensing accuracy. Since the target parameters $\{\theta_{q,l}, d_{q,l}\}$ and environmental factors are unknown at design time, we minimize the expected total loss over possible scenes. Specifically, $\mathcal{S}_l$ denotes the set of randomness sources in the environment in block~$l$, e.g., geometries, reflectivities, Dopplers, and clutters. Thus, the problem is formulated as
\begin{subequations}\label{eq:beam-opt}
\begin{align}
\min_{\mathbf{v}_{m,l},\, P_{m,l}}
\quad & 
\mathbb{E}_{\mathcal{S}_l}
\!\left[
\sum_{q=1}^{Q} 
\ell_{q,l}\!\big(\hat{\theta}_{q,l},\hat{d}_{q,l}\big)
\right]
\label{eq:beam-opt-a}\\
\text{s.t.}\quad 
& \sum_{m = 1}^{M_s} P_{m,l} \le P_{\textrm{Tx}}, \label{eq:beam-opt-c}\\
& \mathbf{v}_{m,l} \in \mathcal{C}_\mathrm{s}\;\; \forall m. \label{eq:beam-opt-d}
\end{align}
\end{subequations}

This problem cannot be optimally solved due to the uncertainty of the target movements and the lack of knowledge of their mobility model. There are well-known classical SP techniques for estimating target parameters from the received echo in radar systems, such as MUSIC \cite{musicref} and ESPRIT \cite{esprit_ref}. However, they rely on fixed array processing assumptions and are not inherently adaptive to temporal variations or uncertain scene conditions. Moreover, in our setup, the design variables $(\mathbf{v}_{m,l},\, P_{m,l})$ shape the transmit signal and subsequently the received echo, and thus, increase the statistical difficulty of the estimation problem in each block. We therefore rely on GAI to estimate target parameters in this dynamic setting. Specifically, we model the conditional distribution of the next-block state $\mathbf{x}_{l+1}$ given the observables available at the BS after block $l$, with the conditioning vector $\mathbf{c}_l$ that collects the transmitter-side observables. In practice, we use the received echo features that are already computed at the BS. These features are noisy and may become less informative as resources become scarce, which motivates a model that can learn and reason under such uncertainty.

\section{Generative Models}

Here, we provide a concise overview of the GAI models that serve as the foundation of our proposed approach. This is intended to offer the necessary background and key principles.

\subsection{DDPM}

DDPMs are iterative generative models that learn to reverse a fixed \emph{forward} noising process. Let us proceed by defining $T_d$ as DDPM timestamps. Hereby, the forward process corrupts clean data $\mathbf{x}_0$ into a sequence $\{\mathbf{x}_t\}_{t=1}^{T_d}$ by Gaussian transitions with a variance schedule $\{\tau_t\}$ such that \cite{DDPM_main}
\begin{align}
q(\mathbf{x}_t \mid \mathbf{x}_{t-1}) &= \mathcal{N}\!\left(\sqrt{1-\tau_t}\,\mathbf{x}_{t-1},\,\tau_t \mathbf{I}\right), \\
q(\mathbf{x}_t \mid \mathbf{x}_0) &= \mathcal{N}\!\left(\sqrt{\bar{\alpha}_t}\,\mathbf{x}_0,\,(1-\bar{\alpha}_t)\mathbf{I}\right),
\end{align}
where $\alpha_t \triangleq 1-\tau_t$ and $\bar{\alpha}_t \triangleq \prod_{s=1}^t \alpha_t$. The model learns \emph{reverse} transitions that denoise step-by-step using
\begin{align}
p_\psi(\mathbf{x}_{t-1} \mid \mathbf{x}_t) &= \mathcal{N}\!\big(\mathbf{x}_{t-1};\,\mu_\psi(\mathbf{x}_t,t),\,\sigma_t^2 \mathbf{I}\big),
\end{align}
where $\sigma_t^2$ can be set to either $\tau_t$ or $\tilde{\tau}_t = \frac{1-\bar{\alpha}_{t-1}}{1-\bar{\alpha}_t}\tau_t$. The mean can be written using a noise-prediction network $\varepsilon_\psi$ as
\begin{align}\label{eq:reverese_DDPM}
\mathbf{x}_{t-1} \;=\; \frac{1}{\sqrt{\alpha_t}}
\!\left(\mathbf{x}_t - \frac{\tau_t}{\sqrt{1-\bar{\alpha}_t}}\,\varepsilon_\psi(\mathbf{x}_t,t)\right) + \sigma_t \mathbf{z},
\end{align}
with $\mathbf{z}\sim\mathcal{N}(\mathbf{0},\mathbf{I})$. Thus, each generation step is stochastic and iteratively denoises $\mathbf{x}_t$ toward $\mathbf{x}_0$. In practice, the denoiser is trained to predict the injected noise at randomly chosen diffusion steps, and sampling starts from $\mathbf{x}_T\!\sim\!\mathcal{N}(\mathbf{0},\mathbf{I})$ and proceeds to $\mathbf{x}_0$.

While DDPM provides a general framework for generative modeling, many applications require conditional generation given side information~$\mathbf{c}$. In a C-DDPM~\cite{C_DDPM_main}, the denoising network~$\varepsilon_\psi$ receives~$\mathbf{c}$ as input so that the reverse process samples from~$p_\psi(\mathbf{x}_0 \mid \mathbf{c})$. A key challenge is to effectively incorporate conditioning while preserving sample quality. Classifier-free guidance~\cite{classifier_free_base} addresses this by enabling conditional sampling without an auxiliary classifier. During training, the model is exposed to both conditional and unconditional data by randomly discarding the conditioning with probability~$p_{\mathrm{drop}}$. This results in two score estimates: the conditional score~$\varepsilon_\psi(\mathbf{x}_t, t, \mathbf{c})$ and the unconditional score~$\varepsilon_\psi(\mathbf{x}_t, t)$. At inference time, these are linearly combined to form a guided score
\begin{align}\label{eq:ddpm_score}
\tilde{\varepsilon}_\psi(\mathbf{x}_t, t, \mathbf{c}) 
= (1+w)\,\varepsilon_\psi(\mathbf{x}_t, t, \mathbf{c}) 
- w\,\varepsilon_\psi(\mathbf{x}_t, t),
\end{align}
where~$w \ge 0$ controls the strength of alignment with~$\mathbf{c}$. The denoising updates then proceed as in the unconditional case, but using~$\tilde{\varepsilon}_\psi$ during sampling. During training, the model is trained to predict the added noise~$\boldsymbol{\epsilon}$ for data samples that have been partially corrupted at a randomly chosen diffusion step~$t \sim \mathrm{Unif}\{{1,\ldots,T_d\}}$ with the objective written as
\begin{align}
\label{eq:ddpm_loss_cf}
L_{\text{simple}}(\psi)
&= \mathbb{E}_{t,\mathbf{x},\mathbf{c},\boldsymbol{\epsilon}}\!
\Big[
\big\|\boldsymbol{\epsilon}
- \varepsilon_{\psi}\big(
\sqrt{\bar{\alpha}_t}\,\mathbf{x}
+ \sqrt{1-\bar{\alpha}_t}\,\boldsymbol{\epsilon},\,
t;\, \mathbf{c}
\big)\big\|^{2}
\Big].
\end{align}

\subsection{VAE}

VAEs are generative models that introduce a latent vector $\mathbf{z} \in \mathbb{R}^{d_z}$ with prior $p(\mathbf{z})=\mathcal{N}(\mathbf{0},\mathbf{I})$ and define a decoder distribution $p_\theta(\mathbf{r}\mid \mathbf{z})$ to generate data samples~\cite{VAE_Tut}. An encoder $q_\phi(\mathbf{z}\mid \mathbf{r})$ approximates the intractable posterior over latents given an observation $\mathbf{r}$, typically as a Gaussian whose mean and variance are predicted by a neural network. Training maximizes the evidence lower bound (ELBO)
\begin{equation}
    \label{eq:elbo}
    L_{\text{ELBO}} =
\mathbb{E}_{\mathbf{z}\sim q_\phi(\mathbf{z}\mid \mathbf{r})}
\!\big[\log p_\theta(\mathbf{r}\mid \mathbf{z})\big]\\
- 
D_{\mathrm{KL}}\!\big(q_\phi(\mathbf{z}\mid \mathbf{r})\,\|\,p(\mathbf{z})\big),
\end{equation}
where the first term encourages accurate reconstruction and the Kullback–Leibler (KL) divergence $D_{\mathrm{KL}}(q\|p)=\mathbb{E}_q[\log(q/p)]$, regularizes the latent distribution toward the prior $p(\mathbf{z})=\mathcal{N}(\mathbf{0},\mathbf{I})$. During inference, new samples are obtained by drawing $\mathbf{z}\sim\mathcal{N}(\mathbf{0},\mathbf{I})$ and decoding with $p_\theta(\mathbf{r}\mid \mathbf{z})$.

\section{DDPM-assisted Framework}

Here, we delve into our proposed framework for multi-target tracking.

\subsection{State Vector}

We represent the per-block target state by stacking a sine–cosine encoding of angles with a logarithmically scaled distance channel. Let $\boldsymbol{\theta}_l$ denote the angles vector and $\mathbf{d}_l$ the distances at block $l$. Then, the DDPM input state is
\begin{equation}
\label{eq:state_pack_sincos}
\mathbf{x}_l \;=\;
\begin{bmatrix}
\sin\boldsymbol{\theta}_l,\
\cos\boldsymbol{\theta}_l,\ 
\rho(\mathbf{d}_l)
\end{bmatrix}^T
\in \mathbb{R}^{3Q},\qquad
\end{equation}
where $    \rho(d) \;=\;
{\log_{10} (d/d_{\min})}/
{\log_{10} (d_{\max}/d_{\min})}$,
and $d_{\min}$ and $d_{\max}$ are the distance bounds. These provide a balanced dynamic range for angles and distances.

\subsection{Conditioning Features}

For the conditioning vector, we leverage the knowledge obtained from the received echo signal $\mathbf{R}_l$. However, $\mathbf{R}_l$ can be a high-dimensional matrix in each block, which makes the model intractable if directly fed into the DDPM conditioner. To cope with this, we leverage a VAE to compress the main features of the echo signal into a low-dimensional latent vector. Let us proceed by using the complex echo $\mathbf{R}_l\in\mathbb{C}^{N_r\times N}$ to form the real two-channel input
\begin{equation}
\bar{\mathbf{R}}_l = \begin{bmatrix}
\Re\{\mathbf{R}_l\}\\[2pt]
\Im\{\mathbf{R}_l\}
\end{bmatrix}
\in\mathbb{R}^{2\times N_r\times N}.
\end{equation}
Then, we apply per-block root mean square (RMS) normalization to obtain
\begin{equation}
\widetilde{\mathbf{R}}_l \;\triangleq\; \sqrt{2N_rN} \bar{\mathbf{R}}_l/||\bar{\mathbf{R}}_l||_F,
\label{eq:vae_rms_norm}
\end{equation}
and encode $\widetilde{\mathbf{R}}_l$ with a VAE by using the posterior mean as latent given by
\begin{align}
q_\phi(\mathbf{z}_l \mid \widetilde{\mathbf{R}}_l) 
&= \mathcal{N}\!\big(\mu_\phi(\widetilde{\mathbf{R}}_l), 
\operatorname{diag}(\sigma^2_\phi(\widetilde{\mathbf{R}}_l))\big), \\
\label{eq:vae_encode}\mathbf{z}_l &= \mu_\phi(\widetilde{\mathbf{R}}_l)\in\mathbb{R}^{d_z}.
\end{align}
Since the normalization removes the absolute echo scale, we also compute a scalar energy feature for the echo given by
\begin{equation}
E_l \;\triangleq\; 10\log_{10}\!\Big(\tfrac{1}{N_rN}\sum_{i=1}^{N_r}\sum_{n=1}^{N}|\mathbf{R}_l[i,n]|^2\Big)\in\mathbb{R}.
\label{eq:echo_energy}
\end{equation}
Finally, we concatenate the VAE latent and the scalar energy into a raw conditioner 
$\mathbf{c}^{\mathrm{raw}}_l = [\mathbf{z}_l^T,\, E_l]^T$.

\subsection{Beam Selection}

One of the main sensing beam design approaches in ISAC systems is to maximize alignment with the target direction. Recall that the C-DDPM takes the conditioner $\mathbf{c}_l$ as input and produces a sample of the next state vector given by $\mathbf{x}_{l+1} \;\sim\; p_\theta(\mathbf{x}_{l+1}\mid \mathbf{c}_l)$,
whose entries correspond to the predicted transmit angles and distances for the next block. For each candidate beam $\mathbf{v}_{m,l} \in \mathcal{C}_\mathrm{s}$, we then evaluate a gain-based score that captures its alignment with these predictions given by
\begin{align}
\label{eq:beam-score}
\mathrm{Score}(\mathbf{v}_{m,l})\;=\; \sum_{q=1}^{Q} \varrho_{q, l+1} \,\norm{\mathbf{a}_t(\hat{\theta}_{q,l+1})^{\mathrm{H}} \mathbf{v}_{m,l}}^2,
\end{align}
where the weight factor $\varrho_{q, l+1}$ is introduced to compensate for the round-trip path-loss attenuation, so that the beam score reflects alignment with the target direction rather than being biased toward closer targets. Then, the top-$M_s$ beams are selected from $\mathcal{C}_\mathrm{s}$, sorted by their scores obtained from \eqref{eq:beam-score}, which directly impacts the quality of the echo in block $l+1$. 

\subsection{Overall framework}

\begin{algorithm}[t]
\caption{C-DDPM-assisted multi-target tracking}
\label{alg:cddpm}
\begin{algorithmic}[1]
\State \textbf{Inputs:} $L,\,N,\,M_s,\,P_{\textrm{Tx}},\,L_{\text{train}},\,K,\,w,\,p_{\text{drop}},\,T_d,$ normalizers, replay $\mathcal{B}$, VAE $\phi$, DDPM $\psi$ 
\State \textbf{Output:} trained VAE $\phi$ and DDPM $\psi$
\State \textbf{Initialization:} choose initial $\mathbf{v}_{m,1}, P_{m, 1}, \forall m$
\For{$l=1$ \textbf{to} $L$}
  \State Transmit $\mathbf{X}_l$ (from $\mathbf{v}_{m,l},\{P_{m,l}\}$), collect echo $\mathbf{R}_l$.
  \State Compute $\widetilde{\mathbf{R}}_l$ and $E_l$ using \eqref{eq:vae_rms_norm} and \eqref{eq:echo_energy}
  \State Encode $\mathbf{z}_l$ using \eqref{eq:vae_encode} and update $\phi$ via \eqref{eq:elbo}
  \State Build $\mathbf{c}^{\mathrm{raw}}_l$ using $(\mathbf{z}_l,E_l)$ and normalize to obtain $\mathbf{c}_l \leftarrow \mathsf{Norm}_c(\mathbf{c}^{\mathrm{raw}}_l)$ \Comment{conditioning for C-DDPM}
  \For{$k=1$ \textbf{to} $K$} \Comment{guided C-DDPM sampling}
     \State Draw $\mathbf{x}^{(k)}_{T_d} \sim \mathcal{N}(\mathbf{0},\mathbf{I}_{3Q})$
     \For{$t = T_d:-1:1$}
        \State Compute $\tilde\varepsilon_\psi(\mathbf{x}^{(k)}_t,t,\mathbf{c}_l)$ using \eqref{eq:ddpm_score}
        \State Compute denoised sample $\mathbf{x}^{(k)}_{t-1}$ using \eqref{eq:reverese_DDPM}
     \EndFor
     \State $\mathbf{x}^{(k)}_0 \leftarrow \mathsf{Norm}_x^{-1}(\mathbf{x}^{(k)}_0)$, then invert \eqref{eq:state_pack_sincos} to obtain $\{\hat{\theta}^{(k)}_{q,l+1},\hat{d}^{(k)}_{q,l+1}\}_{q=1}^Q$.
  \EndFor
  \State Compute scores by averaging \eqref{eq:beam-score} over $K$ samples
  \State Select top $M_s$ beams and set $P_{m,l+1}=P_{\textrm{Tx}}/M_s, \forall m$
  \State Advance scene dynamics to obtain $(\theta_{l+1}, d_{l+1})$.
  \State Form $\mathbf{x}_{l+1}$ using \eqref{eq:state_pack_sincos}
  \State Update EMA normalizers, push $(\mathbf{c}^{\mathrm{raw}}_l,\mathbf{x}_{l+1})$ to $\mathcal{B}$.
  \If{$l \le L_{\text{train}}$} \Comment{per-block DDPM training}
     \State Sample minibatches $(\mathbf{c}^{\mathrm{raw}},\mathbf{x})\!\sim\!\mathcal{B}$ 
     \State Normalize: $\mathbf{c} \leftarrow \mathsf{Norm}_c(\mathbf{c}^{\mathrm{raw}}), \;\; \mathbf{x} \leftarrow \mathsf{Norm}_x(\mathbf{x})$
     \State Draw $t\!\sim\!\mathrm{Unif}\{1{:}T_d\},\;\boldsymbol{\epsilon}\!\sim\!\mathcal{N}(\mathbf{0},\mathbf{I}_{3Q})$
     \State $\mathbf{x}_t=\sqrt{\bar\alpha_t}\,\mathbf{x}+\sqrt{1{-}\bar\alpha_t}\,\boldsymbol{\epsilon}$ 
     \State Drop $\mathbf{c}$ with probability $p_{\text{drop}}$ and update $\psi$ by \eqref{eq:ddpm_loss_cf}
  \EndIf
\EndFor
\end{algorithmic}
\end{algorithm}

Algorithm~\ref{alg:cddpm} summarizes the proposed DDPM-assisted multi-target tracking procedure. Each block begins with probing using the selected beams. Then, the RMS-normalized echo $\widetilde{\mathbf{R}}_l$ is obtained and encoded by a VAE to produce a latent $\mathbf{z}_l$ and an energy feature $E_l$, while the VAE is updated via ELBO steps. The conditioner is normalized with an exponential moving average (EMA)~\cite{EMA_ref} as
\begin{equation}
\mathbf{c}_l = \mathsf{Norm}_c(\mathbf{c}^{\mathrm{raw}}_l) = ({\mathbf{c}^{\mathrm{raw}}_l - \boldsymbol{\mu}_c})/({\boldsymbol{\sigma}_c+\varepsilon}),
\label{eq:cond_norm_sinonly}
\end{equation}
This EMA-based normalization adapts to the non-stationary statistics of echoes across blocks, ensuring stable conditioning for the diffusion model. The EMA-normalized conditioner $\mathbf{c}_l=\mathsf{Norm}_c(\mathbf{c}^{\mathrm{raw}}_l)$ then drives a C-DDPM sampler with classifier-free guidance to draw $K$ trajectories, which are de-normalized and mapped to next-block predictions $\{\hat{\theta}_{q,l+1},\hat{d}_{q,l+1}\}_{q=1}^Q$. In parallel, we apply the same normalization approach by $\mathsf{Norm}_x$ to $\mathbf{x}_l$ (angles and distances) to stabilize training and prediction. Beam scores are computed as the expected amplitude-weighted array gain across $K$ samples, and the top-$M_s$ beams are selected subject to the power budget. For simplicity and fairness, we consider equal power allocation across the selected beams~\cite{isac_beam_alloc_ref}. The environment then advances, we pack $\mathbf{x}_{l+1}$, update normalizers, and push $(\mathbf{c}^{\mathrm{raw}}_l,\mathbf{x}_{l+1})$ to the buffer. When $l \le L_{\mathrm{train}}$, the denoiser is trained per block using minibatches with random diffusion steps, injected noise, and conditioner dropping. The remaining $L-L_{\mathrm{train}}$ blocks are used for inference and evaluation.

\section{Numerical Analysis}\label{subsec:simsetup}

We consider $f=28$~GHz, $N_t=N_r=32$ antennas, $N=64$ slots, and $L=5000$ blocks with $L_{\text{train}} = 400$. Per block, $M_s=8$ probing beams are drawn from a $32$-point DFT codebook. The transmit power is $P_{\mathrm{Tx}}=43$\,{dBm}. The noise power per slot is $\sigma_r^2=-90$\,{dBm}, while
slot and block durations are $T_s = 1$~ms and $T_{\text{blk}} = 64$~ms. We set $r_{\min}=10$~m and $R_{\mathrm{cell}}=50$~m as the minimum and maximum target distance from the BS. We consider $N_c=100$ rank-one static patches, each patch having a fixed angle and distance uniform in $[-\pi/3,\pi/3]$ and $[r_{\min},R_{\mathrm{cell}}]$, respectively. The clutter power is scaled so that the total clutter power is $-55$~{dBm}. Although clutters are static, we include a small Doppler by drawing a per-patch frequency $f'_p \sim \mathcal{N}(0,\sigma_f^2)$ with $\sigma_f=5$. For simplicity and motivated by $\beta_{q,l} \propto 1/d_{q,l}^2$, we consider $\varrho_{q} = \hat{d}_{q}^4$. Although we consider point targets, small RCS fluctuations are included for realistic orientation-dependent or scattering variations. These effects, described later in this section, are abstracted as slow, random changes in the complex coefficient~$\beta_{q,l}$. The DDPM employs a U-Net-based denoising backbone with the 
hidden size as the base channel dimension, the buffer size is $4096$, while the rest of the learning parameters are presented in Table~\ref{tab:learning_params}.

\begin{table}[t]
\centering
\caption{Learning parameters for VAE and DDPM.}
\label{tab:learning_params}
\vspace{-2mm}
\setlength{\tabcolsep}{5pt}
\begin{tabular}{c l c l c}
\toprule
 & \textbf{Parameter} & \textbf{Value} & \textbf{Parameter} & \textbf{Value} \\ 
\midrule
\multirow{2}{*}{\rotatebox[origin=c]{90}{\textbf{VAE}}}
  & Latent dim ($d_z$) & 128 & Hidden size & 256 \\
  & Learning rate & $10^{-3}$ & Epochs/block & 8 \\[2pt]
\midrule
\multirow{4}{*}{\rotatebox[origin=c]{90}{\textbf{DDPM}}}
  & U-Net Hidden size & 512 & Learning rate & $2\times10^{-4}$ \\
  & Diffusion steps ($T_d$) & 200 & Epochs/block & 8 \\
  & Samples ($K$) & 128 & $w$ & 3 \\
  & $\tau_{\text{start}}, \tau_{\text{end}}$ & $10^{-4},10^{-2}$ & $p_{\text{drop}}$ & 0.05 \\
\bottomrule
\end{tabular}
\vspace{-3mm}
\end{table}

\textbf{Mobility:}
Targets follow nearly-constant-velocity dynamics with small Gaussian random fluctuations with zero mean and variance $1$, and occasional small random turns that persist for a few blocks~\cite{mobility_modeling}. Initial angles $\theta_{q,1}$ are on a uniform grid in $[-\pi/3,\pi/3]$, and initial ranges $d_{q,1}\sim\mathcal{U}[r_{\min},R_{\mathrm{cell}}]$. Targets are equally split across the three target types (pedestrian, car, and drone), while the specific parameters are presented in Table~\ref{tab:type-profiles}. Each target’s body orientation is modeled by a smoothly varying heading angle with small random jitters, influencing its RCS and hence the complex coefficient~$\beta_{q,l}$.

\textbf{Reflectivity:} 
We model the effective complex scattering coefficient as
$\beta_{q,l} \;=\; g_{q,l}\,\tilde{\beta}_{q,l}
\;=\; g_{q,l}\,A_{q,l}\,e^{j\phi_{q,l}},
$
where $g_{q,l} = \left({\lambda}/{4\pi d_{q,l}}\right)^{2}$ is the two-way free-space amplitude attenuation, and $\tilde{\beta}_{q,l}$ captures RCS and small-scale scattering. We initialize $\tilde{\beta}_{q,1}\!\sim\!\mathcal{CN}(0,1)$ and set $A_{q,1}\!=\!|\tilde{\beta}_{q,1}|$, $\phi_{q,1}\!=\!\langle\tilde{\beta}_{q,1}\rangle$. Over blocks, $A_{q,l}$ follows a slow log-domain first-order auto-regressive process (AR(1)) around its nominal level with $(0.995,\sigma_{A,\mathrm{dB}})$, rare aspect-gated glints scale $A_{q,l}$~\cite{glint_ref}, and phase evolution is modeled using AR(1) with rare sign flips. We model the clutter coefficient as $\gamma_{p,l}=k_{p,l}\alpha_{p,l}$ with $k_{p,l}$ as two-way free-space attenuation and $\alpha_{p,l}=\rho_c\alpha_{p,l-1}+\sqrt{1-\rho_c^2}\,w_{p,l}$, where $w_{p,l}\sim\mathcal{CN}(0,1)$.


\vspace{-0.4em}
\setlength{\tabcolsep}{3.5pt}
\begin{table}[t]
\centering
\scriptsize
\caption{Type-dependent motion and scattering profiles.}
\label{tab:type-profiles}
\begin{tabular}{@{}l c c c@{}}
\textbf{Parameter} & \textbf{Pedestrian} & \textbf{Car} & \textbf{Drone} \\
\hline
Speed $v_0$ [m/s] & $1.5$ & $15$ & $20$ \\
Random-turn prob, std [deg] & $0.3,\,20$ & $0.1,\,5$ & $0.2,\,10$ \\
Log-amp jitter $\sigma_{A,\text{dB}}$ [dB] & $0.8$ & $0.3$ & $1.0$ \\
Glint prob., strength [dB] & $0.02,\,4$ & $0.01,\,6$ & $0.08,\,10$ \\
Phase AR $\rho_\phi$ & $0.985$ & $0.995$ & $0.975$ \\
Phase-velocity std [deg] & $0.8$ & $0.15$ & $1.2$ \\
Sign flip prob. & $0.02$ & $0.005$ & $0.05$ \\
\hline
\end{tabular}
\end{table}

\vspace{1mm}
\textbf{Benchmarks:} We consider: (i) MUSIC~\cite{musicref} and (ii) ESPRIT~\cite{esprit_ref} for angle estimation, with distances obtained via least-squares; (iii) a convolutional neural network (CNN) regressor that maps the received echo feature to the target states, which consists of three 1D convolutional layers (channels $1{\to}64{\to}128{\to}128$) with ReLU activations, followed by global average pooling and two fully connected layers (sizes $128{\to}128{\to}d_x$); and (iv) a Kalman filter (KF)~\cite{kalman_main}. 

\begin{figure}
    \centering
    \includegraphics[width=0.9\columnwidth]{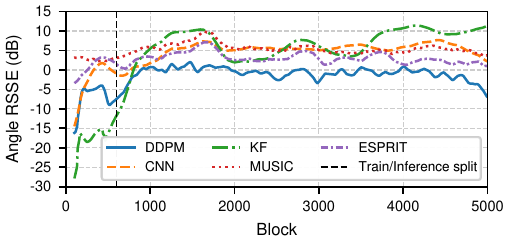}\\
    \includegraphics[width=0.9\columnwidth]{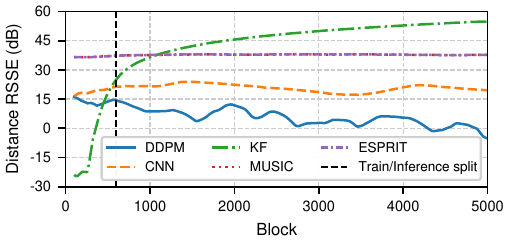}
    \caption{RSSE for (a) estimated angles (top) and (b) estimated distances (bottom) over transmission blocks for $Q = 9$.}
    \label{fig:overblock_errors}
\end{figure}

Fig.~\ref{fig:overblock_errors} reports the root sum square error (RSSE) of estimations over transmission blocks for $Q=9$. During the training phase (with ground-truth feedback), the CNN and KF adapt quickly and initially outperform DDPM. Once inference begins (no ground-truth correction), their regressive updates accumulate drifts, leading to increasing errors. In contrast, the DDPM-assisted tracker leverages its generative nature to adapt to distribution shifts, yielding the lowest error among all baselines during inference.

\begin{figure}
    \centering
    \includegraphics[width=0.9\columnwidth]{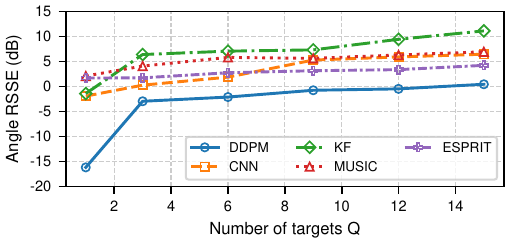}\\
    \includegraphics[width=0.9\columnwidth]{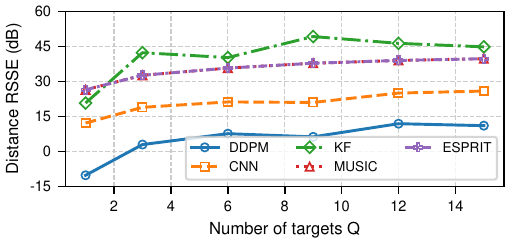}
    \caption{Average RSSE over the inference blocks for (a) angles (top) and (b) distances (bottom) as a function of $Q$.}
    \label{fig:overQ_angle_errors}
\end{figure}

Fig.~\ref{fig:overQ_angle_errors} shows the average RSSE over the inference blocks versus $Q$ across algorithms (no ground truth feedback). The proposed DDPM-assisted framework achieves the best performance with a large margin, both in terms of angle and distance errors. The gains persist as the number of targets increases, indicating DDPM's robustness and effective modeling of the scene’s temporal evolution. It is important to note that ESPRIT and MUSIC employ the same sensing beam design as DDPM, while changing their beam configuration could significantly widen the performance gap in favor of DDPM.


\section{Conclusions}

In this work, we presented an echo-conditioned DDPM–assisted framework for multi-target RF sensing. It learns temporal target dynamics using VAE-based encoded echo and uses classifier-free guidance. The predicted states guide beam selection via the expected amplitude-weighted array gain. Simulations showed consistently lower angle and distance errors than classical SP, filtering, and DL baselines.

\bibliographystyle{ieeetr}
\bibliography{ref_abbv}

\end{document}